\documentclass[]{qrlstm}
\usepackage{graphicx}
\usepackage[hidelinks]{hyperref}
\usepackage{color}
\DeclareUnicodeCharacter{2212}{-}
\AuthorHeaders{Liu, Feng, Feng, Zhu and Liu}

\title{A l\MakeLowercase{earning-based} s\MakeLowercase{tochastic} d\MakeLowercase{riving} m\MakeLowercase{odel} \MakeLowercase{for} a\MakeLowercase{utonomous} v\MakeLowercase{ehicle} t\MakeLowercase{esting}}

\author{%
  \textbf{Lin Liu}\\
  School of Automotive Studies\\
  Tongji University, Shanghai, China, 201804\\
  
  Department of Civil and Environmental Engineering\\
  University of Michigan, Ann Arbor, MI, USA, 48105\\
  linliu0301@163.com\\
  \hfill\break
  \textbf{Shuo Feng, Ph.D., Corresponding Author}\\
  Department of Civil and Environmental Engineering\\
  University of Michigan, Ann Arbor, MI, USA, 48105\\
  fshuo@umich.edu\\
  \hfill\break%
  \textbf{Yiheng Feng, Ph.D.}\\
  University of Michigan Transportation Institute, Ann Arbor, MI, USA, 48105\\
  yhfeng@umich.edu\\
  
  \hfill\break%
  \textbf{Xichan Zhu, Ph.D.}\\School of Automotive Studies\\
  Tongji University, Shanghai, China, 201804\\
  zhuxichan@tongji.edu.cn\\
  \hfill\break%
  \textbf{Henry X. Liu, Ph.D.}\\
  Department of Civil and Environmental Engineering\\
  University of Michigan Transportation Institute\\
  University of Michigan, Ann Arbor, MI, USA, 48105\\
  henryliu@umich.edu
}

\begin{document}
\maketitle
\section{Abstract}
In the simulation-based testing and evaluation of autonomous vehicles (AVs), how background vehicles (BVs) drive directly influences the  AV’s driving behavior and further impacts the testing result. Existing simulation platforms use either pre-determined trajectories or deterministic driving models to model the BVs’ behaviors. However,  pre-determined BV trajectories can not react to the AV's maneuvers, and deterministic models are different from real human drivers due to the lack of stochastic components and errors. Both methods lead to unrealistic traffic scenarios. This paper presents a learning-based stochastic driving model that meets the unique needs of AV testing, i.e. interactive and human-like. The model is built based on the long-short-term-memory (LSTM) architecture. By incorporating the concept of quantile-regression to the loss function of the model, the stochastic behaviors are reproduced without any prior assumption of human drivers.  The model is trained with the large-scale naturalistic driving data (NDD)  from the Safety Pilot Model Deployment (SPMD) project and then compared with a stochastic intelligent driving model (IDM). Analysis of individual trajectories shows that the proposed model can reproduce more similar trajectories to human drivers than IDM. To validate the ability of the proposed model in generating a naturalistic driving environment, traffic simulation experiments are implemented.  The results show that the traffic flow parameters such as speed, range, and headway distribution match closely with the NDD, which is of significant importance for AV testing and evaluation.

\hfill\break%
\noindent\textit{Keywords}: autonomous vehicle; testing and evaluation; stochastic driving behavior; LSTM; quantile regression
\newpage

\section{Introduction}
Testing and evaluation of autonomous vehicles (AVs) become an active research topic in the past few years \cite{kalra2016driving},\cite{zhao2017accelerated},\cite{li2019parallel},\cite{li2020theoretical}. Among three major testing methods (simulation, test track, and on-road) \cite{thorn2018framework}, simulation is the most cost-effective, efficient, and safe method, which attracts significant attention especially in the early development stage of AVs \cite{feng2020testingPartI}, \cite{feng2020testingPartII}, \cite{feng2020testingPartIII}, \cite{feng2020safety}. To evaluate the performance of AV models in a simulation environment, background vehicles (BVs) need to be generated to interact with the AV models in different testing scenarios \cite{feng2021NADE}. To model driver behaviors for the purpose of AV testing, the following features should be included:  

1. Interactive. The BVs should react to the AV’s behavior in real-time.

2. Human-like. The BVs should act like a human driver with stochastic components and errors. For example, different driving styles or mental states of a driver may lead to the variation of driving behavior even in the same traffic environment.

\textcolor{black}{In the past few years, two main methods have been proposed to model BV behaviors for AV testing. In the first method, the BVs behaviors are pre-defined before the testing.  The method is firstly used for testing of advanced driver assistant systems (ADAS) like adaptive cruise control (ACC) and autonomous emergency brake (AEB). The speed profile of the leading vehicle is pre-defined to generate the testing matrix \cite{moon2009design}, \cite{euro2017european}, \cite{arcidiacono2018adas}.} However, the pre-defined BV trajectory cannot reflect the real human driving patterns. Another commonly used approach to apply pre-defined trajectories is to utilize the real-world data collected by vehicles equipped with multiple sensors to replicate the testing scenarios \cite{li2019aads}. Although the behaviors of human drivers can be precisely captured, the main problem of using pre-defined BV trajectories is that the testing is not interactive, because the BVs cannot adjust their maneuvers dynamically based on the AV’s behaviors. 

The second method models the BVs’ behaviors with microscopic traffic flow models, which is an interactive approach. Based on certain driver behavior rules, the motion of each BV at each simulation time step can be updated according to the current traffic state. However, most existing driving behavior models are deterministic such as Newell’s model \cite{newell2002simplified}, Intelligent Driving Model (IDM, \cite{treiber2000congested}), Gipps’ model \cite{gipps1981behavioural}, etc., which cannot capture the real human driver behaviors. Consequently, the AV may pass the test by just ‘remembering’ the experienced scenarios. 

Recently, several studies have been focused on modeling stochastic driving behaviors by adding noise to existing deterministic models. Based on the Newell’s model, Laval et al. \cite{laval2014parsimonious} proposed a stochastic desired acceleration model and added white noise to the driver’s desired acceleration. Treiber et al. \cite{treiber2006understanding} extended the IDM model \cite{treiber2000congested}, optimal-velocity model \cite{bando1995dynamical} and velocity-difference model \cite{jiang2001full} by adding Gaussian white noise to the driver’s desired time gap. However, adding the Gaussian distributed noise to a pre-determined diver model may not reflect real human stochastic driving behaviors, and it is difficult to fit a predefined parametric distribution under different traffic flow conditions (e.g., free flow verse congested). 

 In this paper, we propose a learning-based stochastic driving model to meet the unique needs (i.e., interactive and human-like) for the assessment of AVs. The goal of the model is to generate the action distribution that is consistent with the naturalistic driving data, given current vehicle states. Then by sampling the action from the distribution at each time step, the model can interact with the AV in a stochastic and realistic way. To achieve this goal, the quantile-regression (QR,\cite{koenker2001quantile}) method is incorporated into the learning process. Instead of one single output (e.g., expected acceleration) as commonly designed in existing methods, our method provides a series of outputs, which are designed as the different quantiles of actions. \textcolor{black}{Correspondingly, the pinball loss function \cite{koenker2001quantile} is applied to calculate the loss.} By decreasing the loss function, the output actions learn to be the quantiles, which can fit the naturalistic driving data via the kernel density estimation (KDE,\cite{terrell1992variable}). To further capture the temporal dependency of the behavior model, the long-short-term-memory (LSTM, \cite{zhou2017recurrent}, \cite{wang2019long}) recurrent neural network (RNN) architecture is utilized. \textcolor{black}{The proposed method is referred to as QRLSTM hereafter in this paper.} To validate the effectiveness of the proposed method, a large-scale naturalistic driving database is utilized from the Safety Pilot Model Deployment (SPMD) project \cite{bezzina2014safety}. Simulation results show that the proposed QRLSTM model can represent human driving behaviors at both microscopic and macroscopic levels, which greatly enhances the previous studies by providing a human-like interactive driving environment for AV testing and evaluation.
 
 The new model has two significant advantages. First, it does not apply any assumption of the distribution of human driving error or requires any prior knowledge of human drivers. With the quantile-regression model structure and KDE, the stochasticity of human driving is obtained in a data-driven way. Second, the model has the ability to generate a realistic driving environment, which is of significant value for AV testing and evaluation.   
 
 The rest of this paper is organized as follows. Section 2 formulates the modeling problem and describes the structure of the QRLSTM model, and  Section 3 introduces the model training process. After that, simulation experiments are presented, following by the results and discussions. Conclusions and further research are provided in the final section.

\section{Method}\label{sec:sec2}

\subsection{Problem formulation}
In the introduction section, two features, i.e. interactive and human-like stochasticity, are proposed for the background vehicles to generate a realistic traffic environment for the AV testing. The behavior modeling problem of the background vehicles is formulated as follows.

\textcolor{black}{A microscopic driving model can realize the interaction of the BV with AV as well as other BVs. The goal of a microscopic driver behavior model $f_d$, is to calculate or predict the action (e.g., speed or acceleration) of the BV $\hat{y}_{t+1} $  at the time step $t+1$, given the traffic state $x_t$ at the time step $t$, i.e.
\begin{equation}
  \hat{y}_{t+1}=f_d (x_t),
\end{equation}
where traffic state $x_t$  refers to the dynamic state of BV and all vehicles around it, including AV. In an AV testing simulation, the action of the AV will change the traffic state and BV will calculate the next action accordingly. In this way, the microscopic driving model realizes the interaction between BVs and the AV.}

\textcolor{black}{In the car-following situation, the traffic state $x$ typically only considers the velocity of the BV $v$, the velocity of the AV $v^l$, the range between the BV and the AV $r$, and the range rate between the BV and the AV $rr$, i.e. $x=[v, v^l, r, rr]$.} The action could be either the velocity or acceleration of the BV, denoted as $v$ or $a$. Therefore a car-following model could be written as
\begin{equation}
v_{t+1}=f_d (v_t  ,v^l_t, r_t  ,rr_t )  \text{ or }  a_{t+1}=f_d (v_t  ,v^l_t, r_t  ,rr_t ). 
\end{equation}
For example, the IDM model has the following form:
\begin{equation}
a_{t+1} = a_t[1-(\frac{v_t}{v0})^4-(\frac{s0+v_tT+\frac{v_trr_t}{2\sqrt{ab}}}{r_t})^2],
\end{equation}
where $v0,s0,a,b,T$ are constants \cite{treiber2000congested}. Therefore, the IDM model could be simplified as 
\begin{equation}
a_{t+1}=f_{IDM} (a_t, v_t  , r_t  ,rr_t ).
\end{equation}

However, in actual driving scenarios, human drivers do not behave deterministically, resulting in the need of a stochastic microscopic driving model. With this model, the action of the vehicle might be various even given the same traffic state as input. Several studies have tried to build such a model by adding simple noise (e.g., Gaussian noise) to existing deterministic models. For example, the output of the IDM model is added with a white noise term in \cite{treiber2006understanding}. The modified IDM model could be written as
\begin{equation}
a_{t+1}=f_{IDM} (a_t, v_t  , r_t  ,rr_t ) + \sqrt{Q}\xi_\alpha(t),
\end{equation}
where $Q$ is the fluctuation strength and $\xi_\alpha(t)$ is the white noise. The method of adding white noise to an analytic deterministic model has two disadvantages: the assumption of Gaussian distribution may not reflect real human stochastic driving behaviors, and, the analytic form will disenable the model to fit the changing driving behavior. 

To release the unrealistic assumption of the Gaussian distribution of randomness, a new model structure is proposed in this paper. Instead of calculating the action, the model directly outputs the distribution of action, and the final action is sampled accordingly. The model is defined as
\begin{equation}
    F(\hat{y}_{t+1})=f_d (x_t),
\end{equation}
where $F()$ denotes the distribution function. To achieve such a model structure, we modify the LSTM with quantile-regression loss and kernel density estimation (KDE). The structure of the model is described in detail in the following sections.

\subsection{Model framework}
\begin{figure}[htbp]
    \centering
    \includegraphics[width=12cm]{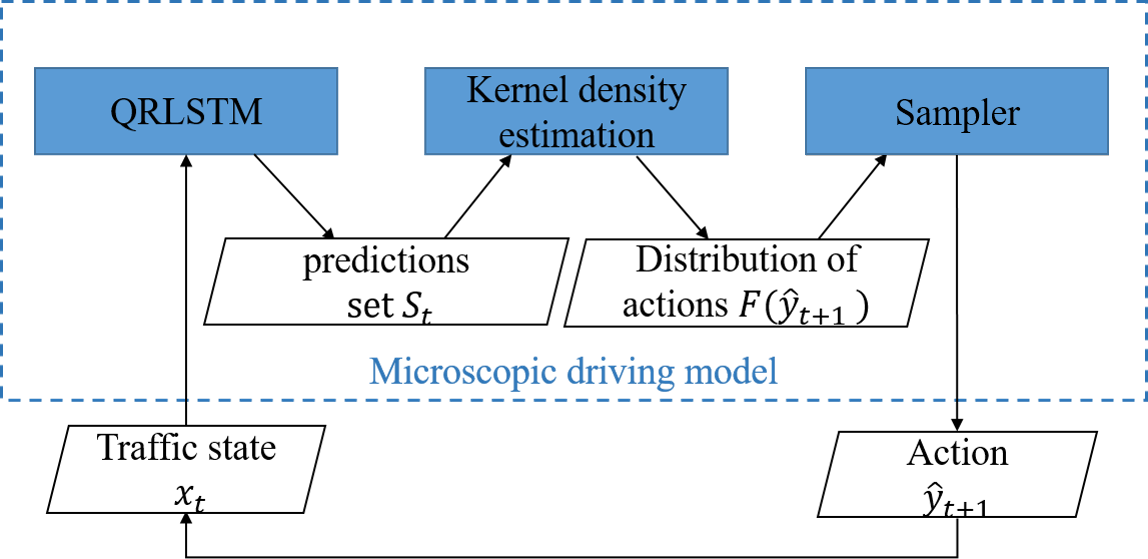}
    \caption{The overall model framework.}
    \label{fig:qrlstm}
\end{figure}

Figure \ref{fig:qrlstm} shows the overall framework of the proposed QRLSTM model, which contains three components: the QRLSTM, KDE, and a sampler.  Given the traffic state $x_t$, the QRLSTM model outputs a set of predicted actions $S_t$ according to the quantile definition $P$. Then, these actions will be used in KDE to estimate the continuous action distribution $F$. Finally, action $\hat{y}_{t+1}$ is sampled from the $F$ distribution. The integration of QRLSTM, KDE, and the sampling process forms a stochastic microscopic driving model. The action $\hat{y}_{t+1}$ will be used to update the traffic state $x_{t+1}$ at time $t+1$, and the loop runs repeatedly.

\subsubsection{LSTM structure}
A significant trend in driver behavior modeling is utilizing machine learning techniques to take advantage of real-world driving data. Neural networks were introduced to model car-following in \cite{hongfei2003develop} and improved by adding human factors by Khodayari et al. \cite{khodayari2012modified}. Different learning-based model structures further improve the modeling performance, such as the deep neural networks model built by Wang et al. \cite{wang2019long} and the reinforcement learning model built by Zhu et al. \cite{zhu2018human}. In \cite{zhou2017recurrent} and \cite{wang2019long}, the RNN  model, which can take the historical state into consideration, shows a better performance in terms of the mean square error in speed prediction. 

\begin{figure}[htbp]
    \centering
    \includegraphics[width=12cm]{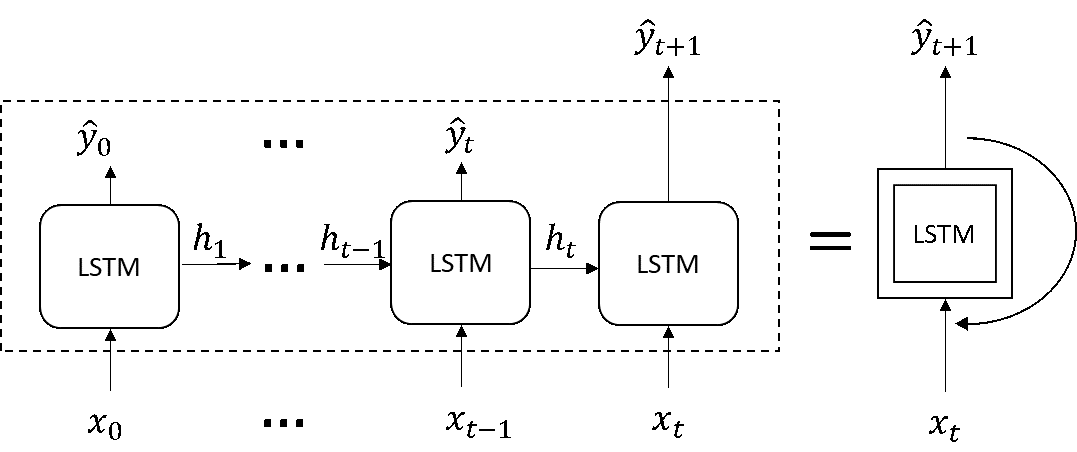}
    \caption{Illustration of the LSTM schema.}
    \label{fig:rnn}
\end{figure}

In this paper, we apply the LSTM neuron network, a widely used neuron network structure, as the base model to calculate the BV action, though our framework is applicable for generic neuron network structures. A simple illustration of LSTM is shown in Figure \ref{fig:rnn}. The detailed neuron structure can be found in \cite{hochreiter1997long}. To calculate $\hat{y}_{t+1}$, LSTM considers both the current input $x_t$ and the hidden state $h_t$, which is calculated based on $x_{t-1}$ and $h_{t-1}$. With this structure, the LSTM can learn both the corresponding output and hidden sequence patterns with sequential training data.

\subsubsection{QRLSTM structure}
\begin{figure}[htbp]
    \centering
    \includegraphics[width=14cm]{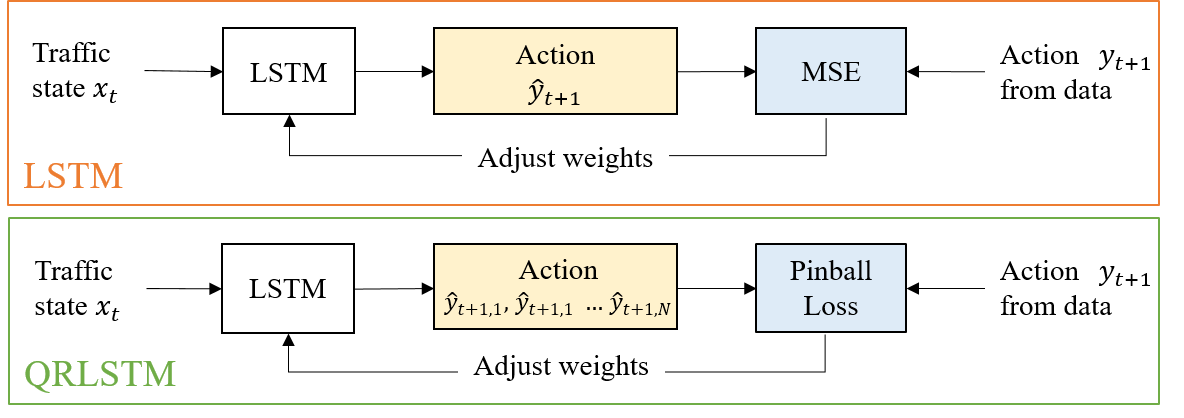}
    \caption{Differences between LSTM and QRLSTM. }
    \label{fig:relstm}
\end{figure}

Due to the randomness and error of human drivers, the action is various even in the same traffic state. Although the driver behavior is recorded as state-action pairs in real-world data, there is an action distribution under a certain traffic state.
However, the LSTM is a deterministic model that outputs one action given a traffic state. As shown in Figure \ref{fig:relstm}, given traffic state $x_t$, LSTM outputs the action $\hat{y}_{t+1}$. The error is then calculated in terms of mean squared error (MSE) for adjusting model weights. Training the model with the MSE cost function will indeed lead the model to estimate the median of actions in a traffic state, which will loss the stochastic information of the real-world data.

To capture the stochasticity of driver behaviors, we propose to apply the concept of quantile-regression \cite{koenker2001quantile}. As shown in Figure \ref{fig:relstm}, the main differences between QRLSTM and LSTM are the output forms of the models and the loss functions used for model training. Specifically, a QRLSTM model outputs a set of $N$ actions $\hat{y}_{t+1,1}, \hat{y}_{t+1,2}... \hat{y}_{t+1,N}$. The error is calculated as the pinball error and used for adjusting the model weights.  By applying the pinball loss as loss function of LSTM, the target of training is changed to estimate action quantiles in a traffic state, instead of the action median. 

The pinball function is designed to calculate the error of a quantile to the real value. Pinball function is defined as
\begin{equation}
    L_{p,y_t}=\left. 
    \begin{cases}
    p(y_t-\hat{y}_{t,p})    &  {if (y_t-\hat{y}_{t,p}) \geq 0}\\
    (p-1)(y_t-\hat{y}_{t,p})   & {if (y_t-\hat{y}_{t,p})< 0}
    \end{cases}
    ,
    \right.
\end{equation}
where $0<p<1$ is the quantile probability, $y_t$ is the observed output from data, $\hat{y}_{(t,p)}$ is the prediction of $p-$quantile, and $L_{(p,y_t)}$ is the loss of the predicted $p-$quantile for $y_t$.
The loss of the QRLSTM model to estimate the $p-$quantile is then defined as
\begin{equation}
    L_p=\frac{1}{N}\sum_{t=2}^{N+1} L_{p,y_t},
\end{equation}
where $N$ is the total number of $y_t$.

Then, QRLSTM is designed to calculate a action matrix corresponding to a set of quantile probabilities, $p\in P$.$P = \{ \frac{1}{|p|-1} , \frac{2}{|P|-1}, ..., 1-\frac{1}{|P|-1}\} $, where, $| P |$is the length of $P$, i.e. the number of quantile probabilities $p$. The Loss function of QRLSTM model is set as
\begin{equation}
    L= \frac{1}{N|P|} \sum_{t=2}^{N+1} \sum_{\forall p\in P}  L_{p,y_t}.
\end{equation}
By training with real-world data, the action set will converge to the action quantiles. 

\subsubsection{Kernel density estimation}

\textcolor{black}{Given a traffic state $x_t$, QRLSTM predicts a $|P| \times 1$  quantile set $S_t= \{ y_{(t,p_1 )},y_{(t,p_2 )},…,y _{(t,p_{|P|})} \} $. To obtain a continuous prediction distribution, KDE \cite{terrell1992variable} is applied.} As a classic non-parametric estimation method, KDE does not require a prior assumption of distribution form, which is suitable for modeling the driver behaviors. The KDE estimation of $S_t$ is calculated by
\begin{equation}
    F(\hat{y}_t) = \frac{1}{B|P|} \sum_{\forall p \in P} K(\frac{\hat{y}_t,p-y_t}{B}),
\end{equation}
where $B>0$ is the bandwidth and $K$ is a kernel function.

\section{Model training }
In this paper, we focus on one of the most common driver behaviors, car-following behavior, to demonstrate the proposed model. To capture the human driving patterns, the QRLSTM model is trained with real driving data from a naturalistic driving study. As in the real traffic environment, different types of drivers have diverse driving behavior styles, a model is trained for individual driver respectively to capture each driver's behaviors.  

\subsection{Data description}
 \textcolor{black}{We adopt the naturalistic driving data (NDD) from the Safety Pilot Model Deployment (SPMD) project \cite{bezzina2014safety} to train and test the proposed model.} With $2,842$ participating vehicles, the SPMD project collected NDD of over $34.9$ million miles in Ann Arbor, Michigan. We utilize the data from 86 vehicles equipped with data acquisition systems (DAS) including MobileEye cameras, which capture vehicle trajectory data (e.g., position, speed, etc.) of the vehicle and its surrounding traffic (e.g., leading vehicles in the same and adjacent lanes) with a frequency of 10Hz. To obtain data in car-following scenarios, the following filtering criteria are applied:
 
1 Road type $= $ highway

2 Speed is larger than $20 m/s$ \textcolor{black}{($72 km/h$)}

3 A leading vehicle is identified by the DAS

Finally, a total number of $24,816$ trajectories are extracted with a total travel time of $1,403,955$ seconds (around 390 hours). Each trajectory lasts for $1$ second to $1,768$ seconds randomly.

\subsection{Model settings}
In the car-following scenario, the traffic state $x$  includes the velocity of the BV $v$, the velocity of the leading vehicle $v^l$, range between the BV and the AV $r$,  and range rate $rr$. The action is the acceleration of the BV $a$. The memory time of the LSTM is set as 10, i.e., the input of the model is the traffic states from the previous 1 seconds, namely, $x_t = \{v_{t-9}, v_{t-8}...v_t;v^l_{t-9}, v^l_{t-8}...v^l_t;r_{t-9}, r_{t-8}...r_t;\\ rr_{t-9}, rr_{t-8}...rr_t\}$. The QRLSTM model has one hidden layer with $32$ LSTM neurons. The quantile probability $P$ is set as $\{0.05, 0.10...0.95\}$. The bandwidth $B$ in KDE is $0.75$, and $K$ is the Gaussian kernel. This model setting is implemented on all QRLSTM models in this paper.

\subsection{Model training}
\begin{figure}[htbp]
    \centering
    \includegraphics[width=11cm]{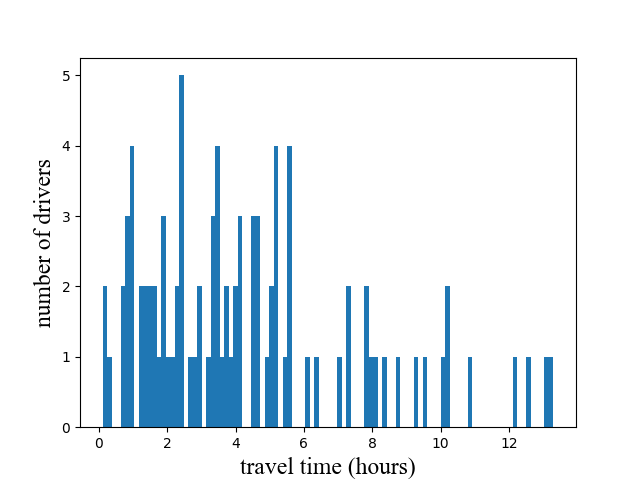}
    \caption{Distribution of total travel time of 86 drivers.}
    \label{fig:travel time}
\end{figure}

\begin{figure}[htbp]
    \centering
    \includegraphics[width=12cm]{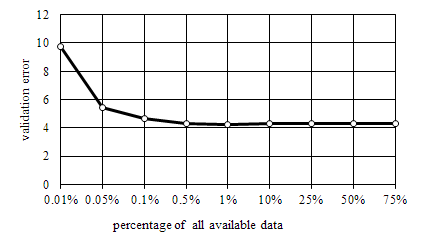}
    \caption{Validation error v.s.  size of training data. }
    \label{fig:percentage}
\end{figure}

Although there are data from 86 drivers available from SPMD, not all of them have enough data to train a QRLSTM model. The total driving time of each driver is shown in Figure \ref{fig:travel time}. There is a large variance in travel time among different drivers. To confirm the required data for the training process, the QRLSTM model is trained with $0.01\%$, $0.05\%$, $0.1\%$, $0.5\%$, $1\%$, $10\%$, $25\%$, $50\%$, and $70\%$ of all available data, respectively, so we can investigate how the validation error of the model changes along with the data size. To calculate the pinball loss as the validation error, $5\%$ of all available data are randomly sampled as validation data. As shown in Figure \ref{fig:percentage}, the validation error drops quickly as the size of the training data grows. The decreasing trend slows down when $0.5\%$ of total data are used in the training, which accounts for about  $7019$ seconds (about 2 hours) driving time. Therefore, 2 hours is set as the threshold of enough data size, and 66 drivers satisfying the condition are modeled individually. 

\begin{figure}[htbp]
    \centering
    \includegraphics[width=12cm]{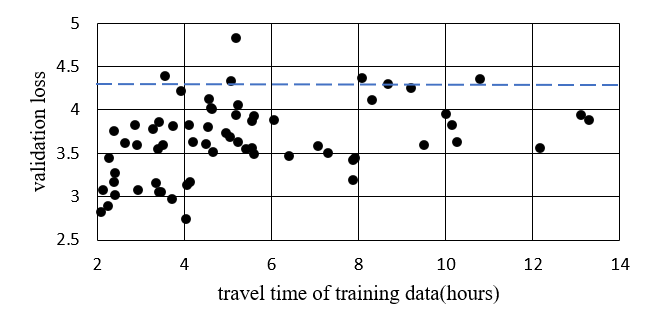}
    \caption{Validation loss.}
    \label{fig:loss}
\end{figure}

The training results of the obtained 66 QRLSTM models are shown in figure \ref{fig:loss}. The average validation loss of individual models is 3.65. The blue dash line stands for the validation loss of the mixed model trained with data of all drivers.  Compared with the validation loss of the mixed model, only 6 individual driver models converge to a higher loss, i.e., $90.91\%$ models outperform the mixed model. The underperformance of the mixed driver against individual models owes to various driving styles and habits of different drivers.

\section{simulation experiments}
\subsection{Experiments Setup}
\begin{figure}[htbp]
    \centering
    \includegraphics[width=16cm]{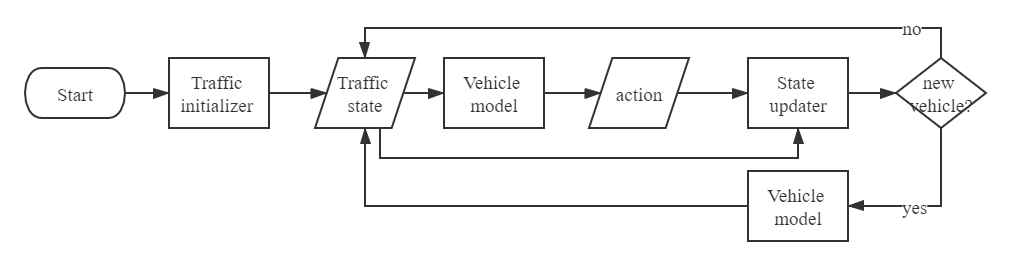}
    \caption{Workflow of mixed driver simulation experiment.}
    \label{fig:simulation}
\end{figure}

\textcolor{black}{To verify the performance of the proposed method, simulation experiments are designed and conducted with Python 3.7, in a work-station equipped with Intel  i7-10700K  CPU  and  16G  RAM.} Since only car-following behavior is considered, a 3-mile single-lane highway is built. The traffic demand varies from 500-2000 vehicles/hour in different experiments. Each experiment lasts for 1 hour and the simulation resolution is 10 Hz. The workflow of the simulation is shown in Figure \ref{fig:simulation}. 

When the simulation starts, the traffic initializer generates the first two vehicles in the network, by randomly sampling the speed of the leading vehicle  $v^l_0$, range $r_0$, and the speed of the following vehicle $v_0$ from the SPMD, and then calculating the corresponding range rate $rr_0$. It forms the traffic state at time step 0, i.e., $x_0={v^l_0,v_0,r_0,rr_0}$. Since the LSTM model structure needs the traffic state of the last 10-time steps as the input, the IDM is used to generate the following 9 time-steps traffic state, to obtain the traffic state of the first 10 time-steps. 

After the initialization, given traffic state at time $t$ for each vehicle model, $s_t={x_{t−9}, x_{t−8}... x_t}$, the vehicle model predicts its action $a_t$.  Then with action $a_t$, the state updater calculates the traffic state for time $t+1$ in the sequential order of vehicle positions. The vehicle generator decides whether a new vehicle should be generated and join the traffic flow. The decision is made by binomial sampling with the probability calculated according to current traffic demand. If a new vehicle is generated, a trained driver model of the 86 drivers will be randomly assigned to the new vehicle according to the travel time distribution as shown in Figure \ref{fig:travel time}.  

\subsection{Result and Discussion}\label{sec:sec3}
\textcolor{black}{As discussed in the introduction, to act as background vehicles in AV testing scenarios, the driver model should interact with the AV, which is already satisfied with the QRLSTM structure and the simulation workflow.} Another requirement is that the model should consist with stochastic human driving behaviors to construct a realistic driving environment. Therefore, performance on both trajectory level and traffic level need to be examined to see whether the vehicle trajectories generated from the proposed model matches the human drivers in the NDD. For comparison, the model proposed by \cite{treiber2006understanding} is replicated as the baseline, which is an extension of the commonly used IDM by adding Gaussian white noise to the driver’s acceleration. The model is defined in Eq (3) and Eq (4)  and the model parameters are also calibrated with the SPMD data set for a fair comparison, shown in Table \ref{tab:IDM}. 

\begin{table}[!ht]
	\caption{Parameters of the modified IDM }\label{tab:IDM}
	\begin{center}
		\begin{tabular}{l l l l}\hline
		    parameter &                  &value\\\hline 
			v0        &  desired velocities & 34.99$m/s$\\
			s0        &  minimum gap        & 1.70$m$ \\
			a        & acceleration       & 0.15$m/s^2$ \\
			b&  comfortable deceleration & 0.66$m/s^2$ \\
			T        & desired time headway  & 0.73$s$ \\
			Q         & fluctuation strength & $0.10m^2/s^3$ \\\hline
		\end{tabular}
	\end{center}
\end{table}

\subsubsection{Trajectory reproducing accuracy}
\begin{figure}[htbp]
    \centering
    \includegraphics[width=12cm]{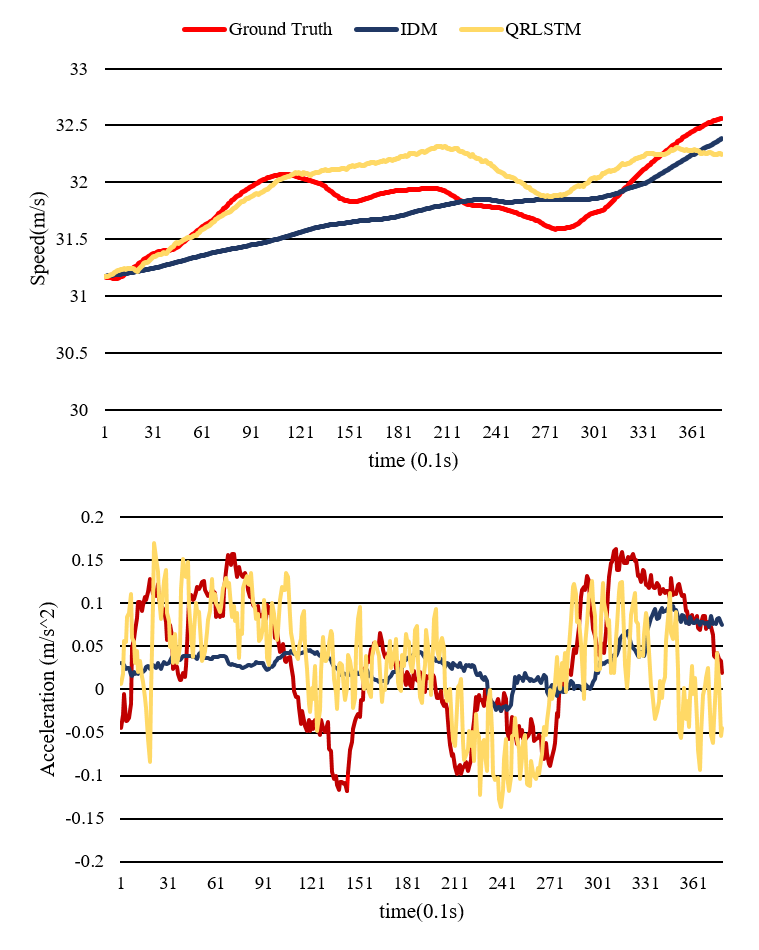}
    \caption{Upper: the continuous speed profile. Lower: the continuous acceleration profile. }
    \label{fig:trajectory}
\end{figure}

Figure \ref{fig:trajectory} shows the comparison at the individual trajectory level among the NDD, QRLSTM, and modified IDM models. The figure represents a continuous action profile of 36s, given the initial condition (i.e., traffic states for the first second). The red, yellow, and blue lines stand for the real vehicle action profile obtained from SPMD, generated by the proposed QRLSTM model, and the modified IDM model, respectively. The upper and lower figures show the action in terms of speed and acceleration respectively. The upper figure shows that the QRLSTM model outperforms the modified IDM model in terms of the speed error, especially for a longer prediction horizon. The lower figure proves that the QRLSTM model is able to capture the frequent variations of acceleration while the modified IDM shows a less sensitive pattern. These frequent changes are indeed from human driving behaviors, which is difficult to be captured by analytical car-following models like IDM. 

\subsubsection{Traffic parameters comparison}
\begin{figure}[htbp]
    \centering
    \includegraphics[width=17cm]{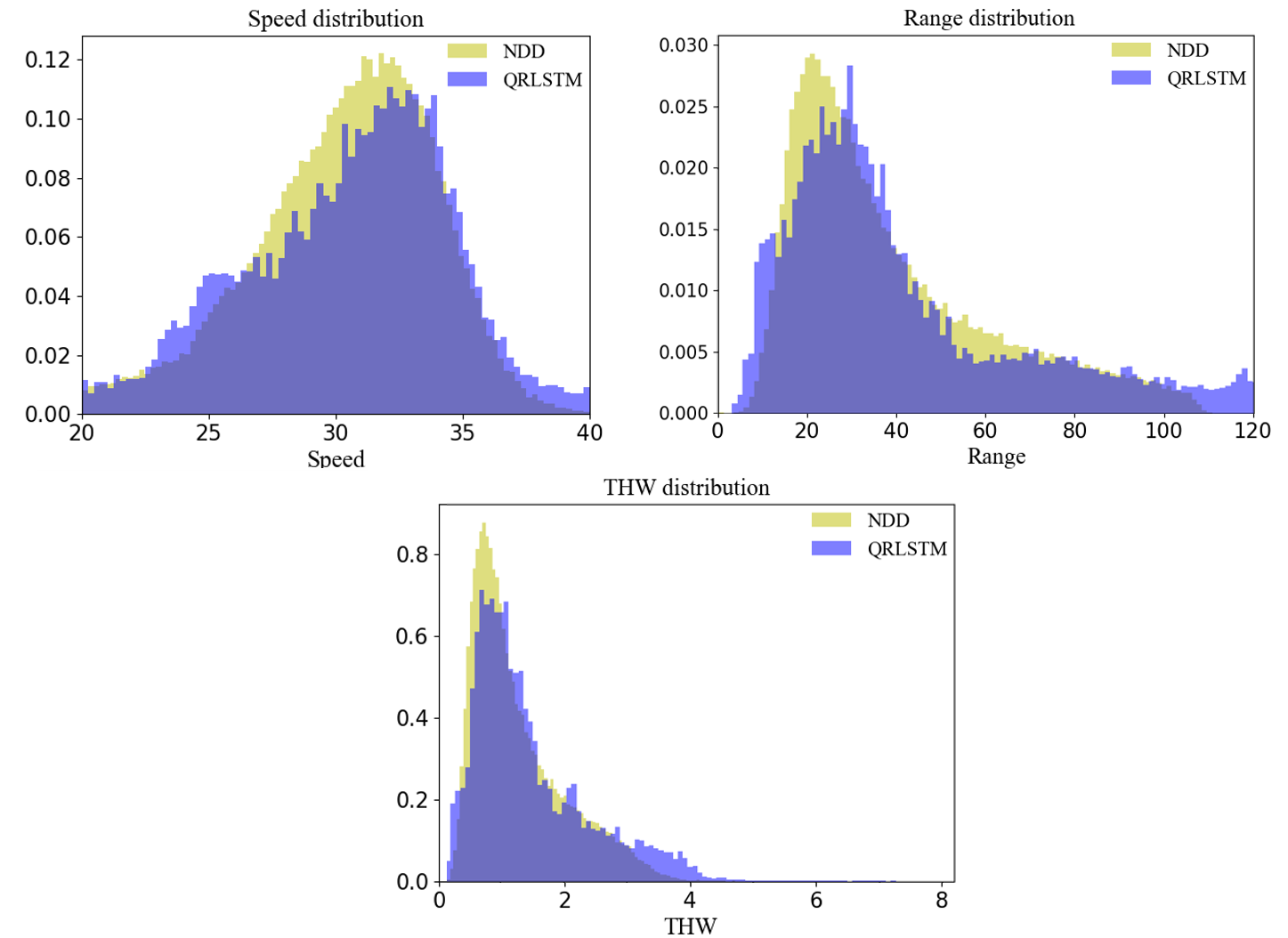}
    \caption{Traffic parameters comparisons between QRLSTM and NDD.}
    \label{fig:result}
\end{figure}

To compare the driving environment generated by the proposed model and the real driving environment,  the distributions of speed, range, and time headway are selected as measurements to describe the traffic flow. A total of 2,386 km of vehicle trajectories are generated by simulation, and 100,000 data points are randomly selected from SPMD to form the distributions. As shown in Fig \ref{fig:result}, the blue distributions represent the proposed QRLSTM model and the yellow distributions represent the real driving data from SPMD. 

Speed and range are two major parameters that can describe the car-following behavior. The upper two sub-figures in Figure \ref{fig:result} present the speed and range distributions respectively. Both distributions of QRLSTM are similar to the SPMD, indicating the similarity between these two traffic environments. Moreover, the Time headway (THW) is also compared, which is critical for the safety test of autonomous vehicles. As shown in the lower figure in Figure \ref{fig:result}, the THW distribution of the QRLSTM model could also capture the trend of the real-world driving environment.

The modified IDM model is also implemented on the same simulation platform as a baseline. To compare the performance, cross-entropy is introduced as a numerical measurement of the similarity between two distributions. The cross-entropy of the distribution $q$ relative to a distribution $p$ within the same sample space $X$ is defined as follows:
\begin{equation}
    H(p,q)=-\sum_{x\in X}p(x)logq(x).
\end{equation}

The smaller the cross-entropy is, the more similar the two distributions are. As shown in Table 2, the cross-entropies
of all three distributions of the QRLSTM model and NDD are significantly smaller than that of the modified IDM and NDD. It means that the distributions of speed, range, and THW of the QRLSTM model are more similar to real-world distributions. Therefore, the proposed QRLSTM model outperforms IDM in terms of generating a more realistic traffic environment.

\begin{table}[!ht]
	\caption{Cross entropy of QRLSTM and IDM with NDD}\label{tab:versions}
	\begin{center}
		\begin{tabular}{l l l l}
			model      & speed    & range   & THW \\\hline
			QRLSTM\& NDD & 0.09500  & 0.44776 & 0.44828 \\
			IDM\& NDD    & 0.11708  & 0.92446 & 0.76247 	\\\hline
		\end{tabular}
	\end{center}
\end{table}

\section{Conclusion and Further Research}
This paper proposed a learning-based stochastic driving model structure for generating a realistic driving environment for AV testing and evaluation purpose. Starting from the well-studied LSTM model structure, the model introduces stochasticity from the quantile-regression-based loss function without any assumptions on the distribution of human driver behavior. The model is trained with real driving data from SPMD and compared with a modified IDM model, showing its superiority over traditional car-following models such as IDM. Microscopic comparison between individual trajectories shows the proposed model is able to capture frequent variations of human driving. Moreover, the comparison at the macroscopic level shows the speed, range, and THW distributions of the proposed model match the NDD distributions well. The results indicate that the traffic environment generated by the proposed model can reflect human driving behaviors. 

This study has several limitations. First, this model is only applied to the car-following scenario in this paper and needs to be extended to scenarios that include lateral vehicle maneuvers such as lane changing and cut-in. \textcolor{black}{Second, as the crash and near-crash scenarios are of great value for AV testing, whether the traffic environment generated with the model can conduct these critical scenarios is another important feature in the naturalistic driving environment modeling. Third, how to integrate physical knowledge into the learning methods deserves more investigation. Recently, the concept of physics regularized machine learning has been proposed for macroscopic traffic flow modeling \cite{yuan2020macroscopic} \cite{shi2021physics}, which is also a promising direction for microscopic behavior modeling.}

\section{Acknowledgement}
The authors would like to thank the US Department of Transportation (USDOT) Region 5 University Transportation Center: Center for Connected and Automated Transportation (CCAT) of the University of Michigan for funding the research. The views presented in this paper are those of the authors alone.

\section{Author Contributions}
The authors confirm contribution to the paper as follows: study
conception and design: Lin Liu, Henry X. Liu, Yiheng Feng, Shuo Feng, Xichan Zhu; data collection: Yiheng Feng, Shuo Feng; analysis and interpretation of results: Lin Liu; draft manuscript preparation: Lin Liu, Shuo Feng, Yiheng Feng, Henry X. Liu. All authors reviewed the results and approved the final version of the manuscript.

\newpage

\bibliographystyle{trb}
\bibliography{trb_template}
\end{document}